\documentclass[groupedaddress]{revtex4}

\usepackage{graphicx}
\usepackage{amssymb}
\usepackage{amsmath}
\usepackage{supertabular}
\usepackage{boxedminipage}
\usepackage{alltt}

\begin{document}

\title{A 15.2 TFlops Simulation of Geodynamo on the Earth Simulator}

\author{Akira Kageyama}
\email{kage@jamstec.go.jp}
\thanks{0-7695-2153-3/04 \$20.00 (c) 2004 IEEE}
\author{Masanori Kameyama}
\author{Satoru Fujihara}
\author{Masaki Yoshida}
\author{Mamoru Hyodo}
\author{Yoshinori Tsuda}
\affiliation{Earth Simulator Center,
Japan Agency for Marine-Earth Science and Technology,
Yokohama 236-0001, Japan}

\begin{abstract}
For realistic geodynamo simulations,
one must solve the magnetohydrodynamic equations
to follow time development of
thermal convection motion of electrically
conducting fluid in a rotating spherical shell.
We have developed a new geodynamo simulation code
by combining the finite difference method with
the recently proposed spherical overset grid called Yin-Yang grid.
We achieved performance of 15.2 Tflops
($46\%$ of theoretical peak performance)
on 4096 processors of the Earth Simulator.
\end{abstract}

\maketitle

\section{Introduction}
The magnetic compass points to the north since the Earth is surrounded
by a dipolar magnetic field.
In the 19th century, Gauss showed that the origin of this
geomagnetic field resides in the Earth's interior.
It is now broadly accepted that the geomagnetic field is generated
by a self-excited electric current in the Earth's core,
which is a double-layered spherical region of radius $r=3500$km.
The inner part of the core,
called the inner core (radius $1200$km), is iron in solid state,
and the outer part, called the outer core
(radius $r$: $1200\hbox{km} \le r \le 3500\hbox{km}$),
is also iron but in liquid state due to the high temperature of the
planetary interior.
The electrical current is generated by
magnetohydrodynamic (MHD) dynamo
action---the energy conversion process from flow energy
into magnetic energy---of the liquid iron in the outer core.
This geodynamo phenomenon is the target of our simulation reported
in this paper.

In the last few decades,
computer simulation has emerged as a central research method
for geodynamo study~\cite{kono_2002}.
 From the beginning of this new and exciting
simulation area,
we have been making important contributions by
combining large scale simulation and
advanced visualization technology:
demonstration of the
strong magnetic field generation by MHD dynamo~\cite{kageyama_1995},
physical mechanism of the dipole field generation~\cite{kageyama_1997},
and spontaneous and repeated
reversals of the dipole moment
(north-south polarity)~\cite{ochi_1999,kageyama_1999,li_2002}.

Recently, we have proposed a new grid system for spherical geometry
based on the overset (or Chimera) grid methodology.
We have applied this new spherical grid
to geodynamo simulation.
In this paper, we report the development of this
new geodynamo simulation code and its performance on the
Earth Simulator~\cite{shingu_2002,yokokawa_2002,habata_2003},
of Japan Agency for Marine-Earth Science and Technology (JAMSTEC), Japan.

\section{Yin-Yang grid}
Since the Earth's outer core  is a spherical shell region
between the inner core and the mantle,
we need a numerically efficient spatial discretization method for a
spherical shell geometry in order to achieve high performance on
massively parallel computers.

In our previous geodynamo simulations,
we used a finite difference method
based on the latitude-longitude grid
in spherical polar coordinates
with the radius $r$ ($r_i\le r\le r_o$),
colatitude $\theta$ ($0\le \theta\le \pi$),
and the longitude $\phi$ ($-\pi < 0\le \phi \le \pi$).
Due to the existence of the
coordinate singularity and grid convergence
near the poles of the latitude-longitude grid,
we had to take special care at the poles
and this inevitably degraded the numerical efficiency and
the performance of our previous dynamo code.

Since the finite difference method is suitable for
massively parallel computers,
especially massively parallel vector supercomputers
like the Earth Simulator,
we have decided to further exploit the
possibilities of the finite difference method
in spherical geometries, and
to explore a new spherical grid system
as the base of the spherical finite difference method.

There is no grid mesh that is orthogonal all over a
spherical surface and, at the same time, free of
coordinate singularity or grid convergence problems.
So, we decompose the spherical surface into subregions.
The decomposition, or dissection, enables us to cover each subregion
by a grid system
that is individually orthogonal and singularity-free.

The dissection of the computational domain generates
internal borders between the subregions.
In the overset grid approach~\cite{chesshire_1990},
the subdomains are permitted to
partially overlap one another on their borders.
The overset grid is also called an
overlaid grid,
or composite overlapping grid,
or Chimera grid~\cite{steger_1983}.
It is now one of the most important grid techniques
in computational aerodynamics.

Recently, we have proposed a new overset grid for spherical
geometry~\cite{kageyama_2004}.
The grid is named ``Yin-Yang grid'' after the symbol
for yin and yang in the Chinese philosophy of complementarity.
The Yin-Yang grid is composed of two identical and complemental
component grids.
Compared with other possible spherical overset grids,
the Yin-Yang grid is simple in its geometry as well as
metric tensors.
A remarkable feature of this overset grid is that
the two identical component grids are combined in a complemental way
with a special symmetry.
The Yin-Yang grid has already been
applied to a mantle convection simulation
in a spherical shell
geometry with detailed benchmark tests~\cite{yoshida_2004}.
The Yin-Yang grid has also been applied to simulations of
the global circulation of the atmosphere, ocean, and their
coupled system \cite{takahashi_2004,xindong_2004,komine_2004,ohdaira_2004,hirai_2004}.

\begin{figure}[t]
  \begin{center}
   \includegraphics[width=0.99\textwidth]{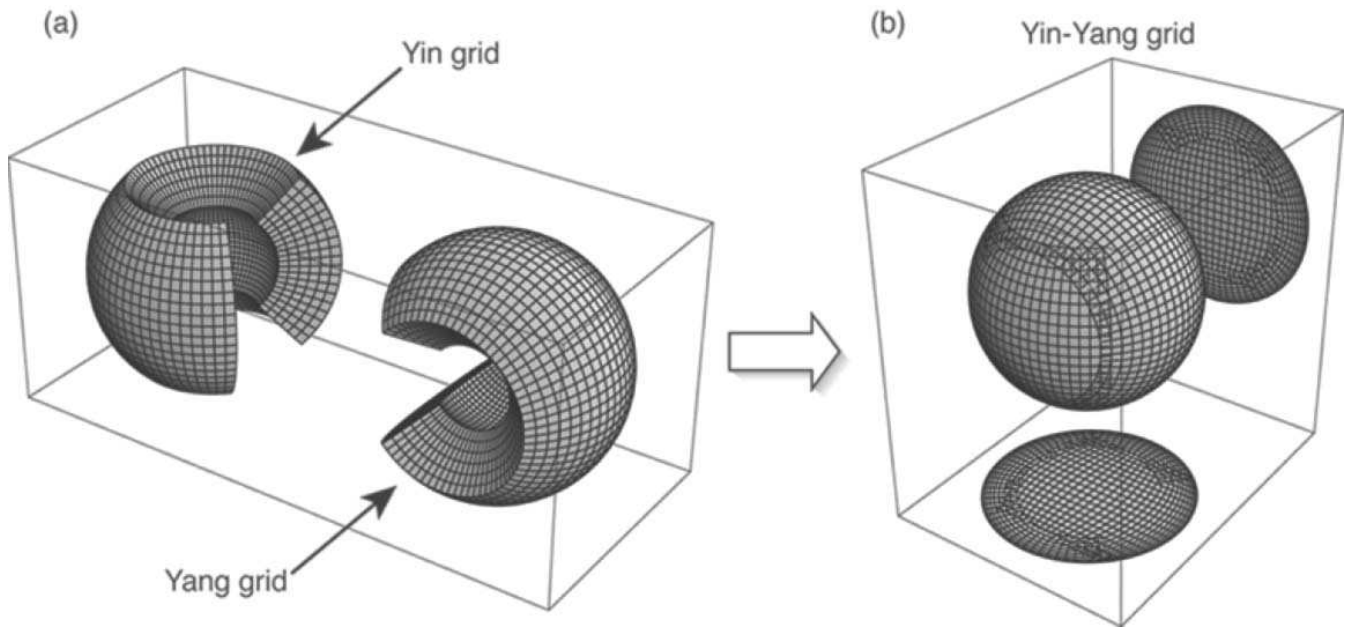}
  \end{center}
  \caption{Basic Yin-Yang grid.
The Yin grid and Yang grid
are combined to
cover a spherical surface with partial overlap.
}
  \label{fig:basicYinYang}
\end{figure}
There are several varieties of Yin-Yang grid~\cite{kageyama_2004}.
The most basic type, adopted in the geodynamo simulation
reported in this paper,
is shown in Fig.~\ref{fig:basicYinYang}.
It has two components---Yin and Yang---that
are geometrically identical
(exactly the same shape and size);
see Fig.~\ref{fig:basicYinYang}(a).
We call the two component grids the ``Yin grid'' (or \textit{n}-grid)
and the ``Yang grid'' (or \textit{e}-grid).
They are combined to cover a spherical shell with partial overlap
on their borders as shown in Fig.~\ref{fig:basicYinYang}(b).
Each component grid is, in fact,
a part of the latitude-longitude grid on a spherical surface.
(It is piled up in the radial direction).
A component grid, say the Yin grid,
is defined as a part of low-latitude
region of the usual latitude-longitude grid:
$90^\circ$ around the equator in latitude (between
$45^\circ$N and $45^\circ$S),
and $270^\circ$ in longitude.
Another component grid, the Yang grid, is defined in the same way
but in different spherical coordinates.
The virtual north-south axis of the Yang grid
is located on the equator of the Yin grid's coordinates.
The relation between Yin coordinates and Yang coordinates
is written in Cartesian coordinates by
\begin{equation}
  (x^e,y^e,z^e) = (-x^n,z^n,y^n), \ \ \ \hbox{or} \ \ \
  (x^n,y^n,z^n) = (-x^e,z^e,y^e),   \label{eq:a05}
\end{equation}
where $(x^n,y^n,z^n)$ is the Yin grid's Cartesian coordinates
and $(x^e,y^e,z^e)$ is Yang grid's.
Note that the forward and inverse transformations
between Yin and Yang are given in the same form,
which is a reflection of the complementary relation between the
Yin and Yang grids.

It should be stressed that
the Yin grid and Yang grid are identical in shape,
geometry, grid, and metric tensors.
This means that all subroutines
designed for, say the Yin grid, can be used for
the Yang grid with no modification.
In addition to this geometrical symmetry,
the Yin grid and Yang grid are in complementary
coordinates relative to each other as indicated by eq.~(\ref{eq:a05}).
This means that any interaction or communication from
a grid point on Yin to a grid point on Yang
is exactly the same as that from Yang to Yin.
This fact makes the Yin-Yang-based computer
code very concise and efficient.

Another advantage of the Yin-Yang grid
resides in the fact that the component grid (Yin or Yang) is
nothing but a (part of) latitude-longitude grid.
This means that
we can directly deal with the equations to be solved with
the vector form in the usual spherical polar coordinates;
components of a flow vector $\mathbf{v}$, for example, can be
written as $\{v_r, v_\theta, v_\phi\}$ in the program.
The analytical forms of metric tensors such
as for Laplacian in spherical coordinates are well known.
Since we can directly code
the basic equations as they are formulated in
spherical coordinates,
we can make use of various resources of
mathematical formulas, program libraries, and tools
that have been developed in spherical polar coordinates.

Following the general overset methodology~\cite{chesshire_1990},
interpolations are applied on the boundary of each component grid
to set the boundary values, or internal boundary conditions.
When one sees the component grid of the basic Yin-Yang grid
shown in Fig.~\ref{fig:basicYinYang} in the Mercator projection,
it is a rectangle; the four corners intrude most into the
other component grid.
Even if the grid mesh is taken to be infinitesimally small, the
overlapping area has still non-zero ratio of about $6\%$
of the whole spherical surface.
We obtain two solutions---one from the Yin side and other from
the Yang side---in the overlapped area.
This might be a slight ($6\%$) waste of computational time,
but the ``double solution'' causes no problem in actual
calculations.
The difference between the two solutions is within
the discretization error that is omnipresent on the sphere
in any case.
The difference between the solutions is so small that
we do not need to ``blend'' the double solutions there
in the post processing for the
data visualization;
we just choose one of the two solutions
and the resulting visualization shows smooth pictures.
There is no indication of the
internal border between the Yin and Yang grids
(e.g., Fig.~\ref{fig:columns} below).

If one still desires to minimize
the overlapped area, it is possible by modifying the
component grid's shape from the rectangle.
It is obvious that a Yin-Yang grid with a minimum overlap
region can be constructed by a division, or dissection,
using a closed curve on a sphere that cuts
the sphere into two identical parts.
There are infinite number of such dissections of a sphere.
Two examples of them---``baseball type dissection'' and
``cube type dissection'' can be found in \cite{kageyama_2004}.

\section{Basic equations and numerical methods}
Since the physical model for the geodynamo simulation
is the same as our previous one and
described in our papers~\citep[e.g.,][]{kageyama_1995},
here we describe it only briefly.
We consider a spherical shell vessel bounded by two concentric
spheres.
The inner sphere of radius $r=r_i$ denotes the
inner core and the outer sphere of $r=r_o$ denotes the
core-mantle boundary.
An electrically conducting fluid is confined in this shell region.
Both the inner and outer spherical boundaries rotate with a constant
angular velocity ${\bf \Omega}$.
We use a rotating frame of reference with the same angular velocity.
There is a central gravity force in the direction of the center of
the spheres.
The temperatures of both the inner and outer spheres are fixed; hot
(inner) and cold (outer).
When the temperature difference is sufficiently large, a convection
motion starts when a random temperature perturbation is imposed at
the beginning of the calculation.
At the same time an infinitesimally small,
random ``seed'' of the magnetic field is given.

The system is described by the following normalized MHD equations:
\begin{equation} \label{eq:continuity}
     \frac{\partial \rho}{\partial t} = - \nabla \cdot \mathbf{f},
\end{equation}
\begin{equation} \label{eq:motion}
    \frac{\partial \mathbf{f}}{\partial t}
                   = -\nabla \cdot (\mathbf{v}\mathbf{f})
                     - \nabla p
       + {\bf j} \times {\bf B}
                     + \rho {\bf g}
                     + 2 \rho {\bf v}\times{\bf \Omega}
                     + \mu (   \nabla^2 {\bf v}
                     + \frac{1}{3} \nabla (\nabla \cdot {\bf v} ) ),
\end{equation}
\begin{equation} \label{eq:press}
         \frac{\partial p}{\partial t}
                = - \mathbf{v}\cdot \nabla p
                 - \gamma p \nabla \cdot {\bf v}
	              + (\gamma-1) K \nabla^2 T
		             +   (\gamma-1)\eta {\bf j}^2
			            +   (\gamma-1)\Phi,
\end{equation}
\begin{equation} \label{eq:induction}
   \frac{\partial {\bf A}}{\partial t} = - {\bf E},
\end{equation}
with
\begin{eqnarray}
\nonumber
                &    p = \rho T, \hspace{3em}
                    {\bf B} =  \nabla \times {\bf A}, \hspace{3em}
                     {\bf j} =  \nabla \times {\bf B}, \hspace{3em}
          {\bf E} = - {\bf v} \times {\bf B} + \eta {\bf j},        \\
                &
                    {\bf g} = - g_0/r^2 {\bf \hat{r}}, \hspace{3em}
              \Phi =
                2 \mu \left( e_{ij} e_{ij}
	 - \frac{1}{3} (\nabla \cdot {\bf v})^2 \right), \hspace{3em}
       e_{ij} = \frac{1}{2} \left( \frac{\partial v_i}{\partial x_j}
       + \frac{\partial v_j}{\partial x_i}
          \right).
\end{eqnarray}
Here the mass density $\rho$, pressure $p$,
mass flux density $\mathbf{f}$,
magnetic field's vector potential $\mathbf{A}$ are
the basic variables in the simulation.
Other quantities;
magnetic field ${\bf B}$,
electric current density ${\bf j}$,
and electric field ${\bf E}$ are treated as subsidiary fields.
The ratio of the specific heat $\gamma$, viscosity $\mu$,
thermal conductivity $K$ and electrical resistivity $\eta$ are
assumed to be constant.
The vector ${\bf g}$ is the gravity acceleration and ${\bf \hat{r}}$ is the
radial unit vector; $g_0$ is a constant.
We normalize the quantities as follows:
The radius of the outer sphere $r_o = 1$; the temperature of the
outer sphere $T(r_o)$ = 1; and the mass density at the outer sphere
$\rho(r_o) = 1$.

The spatial derivatives in the above equations are
discretized by second-order central finite differences
in spherical coordinates:
($r$, $\theta$, $\phi$).
The fourth-order Runge-Kutta method is used for
the temporal integration.

There are six free
parameters---including three dissipation
constants; viscosity $\mu$, thermal conductivity $K$,
and electrical resistivity $\eta$---in this MHD system.
We basically set the same parameter values
adopted in our previous simulation
that showed repeated dipole reversals~\cite{li_2002}
except for the three dissipation constants;
we set each of them 10 times smaller than in the previous simulation.
This means that the fluid's Reynolds number
and magnetic Reynolds number are both 10 times larger,
the Rayleigh number---an index of
the vigor of thermal convection---is 100 times larger, $3\times 10^6$,
and the Ekman number is $2\times 10^{-5}$.
Thus we can achieve more turbulent, and therefore more realistic
simulations compared with our previous simulations.

\section{Parallel Programming and Performance on the Earth Simulator}
We have developed a new geodynamo simulation code
(hereafter ``yycore'' code)
for the Earth Simulator
by converting our previous geodynamo code, which was
based on the traditional latitude-longitude grid,
into the Yin-Yang grid.
We use the Fortran90 language and the MPI library.
We make use of the module unit of Fortran90
to keep concise and neat program structure.
All subroutines and functions are module subprograms.
(There are no external subprograms.)
In order to realize data hiding and
encapsulation in the module level,
we set the default attribute of module data and
subprograms be \texttt{private}.
The yycore code has about 6,200 lines of source (including comments).
There are 12 modules and 1 main program.
The total number of module-subprograms are 153.
Among them, 48 subprograms are public (global scope).

We have found that
the code conversion from our previous latitude-longitude-based code
into the new Yin-Yang-based code is straightforward and rather easy.
This is because most of the Yin-Yang grid code
shares source lines with the latitude-longitude grid code:
Our previous geodynamo code was basically a finite-difference
MHD solver on spherical coordinates with a full span
of colatitude $(0\le \theta \le \pi)$
and longitude $(-\pi < \phi \le \pi)$;
on the other hand,
the Yin-Yang grid code (yycore) is also a finite-difference
MHD solver on the spherical coordinates, but with just the
smaller span of colatitude ($\pi/4 \le \theta \le 3\pi/4$)
and longitude ($-3\pi/4 \le \phi \le 3\pi/4$).
The major difference is the new boundary condition
(interpolation)
for the communication between Yin grid and Yang grid.

Our experience with the rapid and easy conversion
from latitude-longitude code into Yin-Yang code
should be encouraging for others who have already developed
codes that are based on latitude-longitude
grids in the spherical coordinates,
and who are bothered by numerical problems and inefficiency
caused by the pole singularity.
We would like to suggest that they try the Yin-Yang grid.

Since the Yin grid and Yang grid are identical,
dividing the whole computational domain into a Yin grid part
and a Yang grid part
is not only natural but also efficient for parallel processing.
In addition to this Yin-and-Yang division,
further domain decomposition within each grid
is applied to achieve high performance
on the Earth Simulator, a massively parallel computer.

The Earth Simulator, whose hardware specifications are
summarized in Table~I has
three different levels of parallelization:
vector processing in each arithmetic processor (AP);
shared-memory parallelization by 8 APs in each processor node (PN);
and distributed-memory parallelization by PNs.

In our yycore code,
we apply vectorization in the radial dimension
of the three-dimensional (3D) arrays for physical variables.
The radial grid size is 255 or 511,
which is just below the size (or doubled size) of
the vector register of
the Earth Simulator (256) to avoid
bank conflicts in the memory~\cite{shingu_2002}.

We use MPI both for the inter-node (distributed memory)
parallel processing
and for the intra-node (shared memory) parallel processing.
This approach is called ``flat-MPI'' parallelization.

As we mentioned above,
we first divide the
whole computational domain into two identical
parts---here we call them ``panels''---that
correspond to the Yin grid's domain and the Yang grid's domain
in Fig.~\ref{fig:basicYinYang}(a).
In the yycore code, \texttt{MPI\_COMM\_SPLIT} is invoked to
split the whole processes into two groups.
(The total number of processes is even.)
The world communicator is stored in a variable named
\texttt{gRunner\%world\%communicator}, where
\texttt{gRunner} is a nested derived type.

For further parallelization within a panel,
there are several choices for the method of domain decomposition.
Here we choose two-dimensional decomposition
in the horizontal space, colatitude $\theta$ and longitude $\phi$,
in each panel.
For this purpose,
we call \texttt{MPI\_CART\_CREATE} to make a two-dimensional
process array with optimized rank order.

For the intra-panel communication,
\texttt{MPI\_SEND} and \texttt{MPI\_IRECV}
are called between nearest neighbor processes.
Each process has four neighbors (north, east, south, and west).
The rank numbers for the neighbors, obtained by invoking
\texttt{MPI\_CART\_SHIFT} with the panel's communicator
\texttt{gRunner\%panel\%communicator}, are also stored
in \texttt{gRunner}.

Communication between two groups (Yin and Yang)
is required for the overset interpolation.
This communication is implemented by
\texttt{MPI\_SEND} and \texttt{MPI\_IRECV}
under \texttt{gRunner\%world\%communicator}.

\noindent
\begin{table}
\begin{tabular}{|r|r|}
\hline
Peak performance of arithmetic processor (AP) & 8 Gflops \\ \hline
Number of AP in a processor node (PN) & 8 \\ \hline
Total number of PN & 640 \\ \hline
Total number of AP & $8\hbox{ AP}\times 640\hbox{ PN}=5120$ \\ \hline
Shared memory size of PN & 16 GB \\ \hline
Total peak performance &
          $8 \hbox{ Gflops}\times 5120\hbox{ AP}=40 \hbox{Tflops}$ \\ \hline
Total main memory & 10TB\\ \hline
Inter-node data transfer rate & 12.3 GB/s $\times$ 2 \\ \hline
\end{tabular}
\label{tab:00}
\caption{Specifications of the Earth Simulator.}
\end{table}

The best performance of yycore code with this flat MPI
parallelization is $15.2$ Tflops.
This number was obtained from the hardware counter for floating point
operations on the Earth Simulator, which is
automatically reported by setting the
environment variable \texttt{MPIPROGINF}.
This performance
is achieved by $4096$ processors (512 nodes) with the total grid
size of $511(\hbox{radial})
\times 514(\hbox{latitudinal})
\times 1538(\hbox{longitudinal})
\times 2(\hbox{Yin and Yang})$.
Since the theoretical peak performance of $4096$ processors
is $4096\times 8\hbox{ Gflops}=32.8 \hbox{Tflops}$,
we have achieved $46\%$ of peak performance in this case.
The average vector length is $251.6$,
and the vector operation ratio is $99\%$.
These data can be seen in
the output of \texttt{MPIPROGINF} shown in List~I.
The high performance of the yycore code is
a direct consequence of
the simple and symmetric configuration design
of the Yin-Yang grid:
It makes it possible to minimize the
communication time ($10\%$) between the
processes in the horizontal directions,
and enables optimum vector processing (with $99\%$ of operation ratio)
in the radial direction in each process.

\vspace{0.5em}
\noindent
{\footnotesize
\begin{center}
\begin{boxedminipage}[h]{0.95\textwidth}
\begin{alltt}
     MPI Program Information:
     ========================
     Note: It is measured from MPI_Init till MPI_Finalize.
           [U,R] specifies the Universe and the Process Rank in the Universe.
     
     Global Data of 4096 processes:          Min [U,R]             Max [U,R]         Average
     =============================
     
     Real   Time (sec)           :      452.157 [0,2623]      454.266 [0,680]       453.457
     User   Time (sec)           :      441.499 [0,1741]      447.001 [0,7]         443.220
     System Time (sec)           :        4.232 [0,455]         5.499 [0,9]           4.498
     Vector Time (sec)           :      321.969 [0,245]       380.540 [0,2138]      351.678
     Instruction Count           :  45628623153 [0,3152]  48004139033 [0,2]     46732455581
     Vector Instruction Count    :  13505506552 [0,855]   14116925576 [0,2178]  13758270302
     Vector Element Count        : 3395555533860 [0,1904] 3553256715628 [0,2178] 3461109543510
     FLOP Count                  : 1641123079258 [0,52]   1668645286059 [0,2178] 1642792822350
     MOPS                        :     7706.674 [0,7]        8102.270 [0,3938]     7883.403
     MFLOPS                      :     3671.412 [0,7]        3769.202 [0,1890]     3706.499
     Average Vector Length       :      250.582 [0,1120]      252.856 [0,4068]      251.564
     Vector Operation Ratio (%)  :       99.000 [0,7]          99.105 [0,3135]       99.056
     Memory size used (MB)       :     1042.944 [0,0]        1122.913 [0,9]        1106.882
     
     
     Overall Data:
     =============
     
     Real   Time (sec)           :      454.266
     User   Time (sec)           :  1815428.791
     System Time (sec)           :    18425.031
     Vector Time (sec)           :  1440471.743
     GOPS   (rel. to User Time)  :    32290.442
     GFLOPS (rel. to User Time)  :    15181.807 \ \ \ \ \ \ \  <--- 15.2 TFlops
     Memory size used (GB)       :     4427.528
\end{alltt}
\end{boxedminipage}
\end{center}
}
\begin{center}
\noindent
List~1: An example of \texttt{MPIPROGINF} output.
\end{center}

\vspace{0.5em}
To measure the performance dependency on the total number of processes,
and the problem size (grid number),
we executed several runs that are summarized in Table~II.

\begin{table}
\begin{tabular}{|r|r|r|r|}
\hline
processors & grid points & Tflops & efficiency \\
\hline
4096 & $511\times 514\times 1538\times 2$ & 15.2  & 46\% \\
3888 & $511\times 514\times 1538\times 2$ & 13.8  & 44\% \\
3888 & $255\times 514\times 1538\times 2$ & 12.1  & 39\% \\
2560 & $511\times 514\times 1538\times 2$ & 10.3  & 50\% \\
2560 & $255\times 514\times 1538\times 2$ &  9.17 & 45\% \\
1200 & $255\times 514\times 1538\times 2$ &  5.40 & 56\% \\
\hline
\end{tabular}
\label{tab:01}
\caption{Performance achieved by the yycore
code---a finite difference geodynamo simulation
program based on the Yin-Yang grid---
on the Earth simulator.}
\end{table}

Generally, flat MPI parallelization requires
a larger problem size to achieve the same level of
performance efficiency
compared to the hybrid parallelization
(e.g., MPI for inter-node and microtasking for intra-node parallelization)
on the Earth Simulator \cite{nakajima_2002}.
Since one Earth Simulator node has 8 APs (see TABLE~I),
the flat MPI method generates 8 times as many MPI processes
as hybrid parallelization.
However, in our yycore code with flat MPI,
high performance could be achieved
with relatively low numbers of mesh size:
$15.2$ Tflops/512 PN
with $511\times 514\times 1538\times 2=8\times 10^8$ grid points;
$12.1$ Tflops/486 PN
with
$255\times 514\times 1538\times 2=4\times 10^8$ grid points.
This can be compared with other flat MPI simulations
on the Earth Simulator as
reported in SC2002 and SC2003 (see Table~III).
Sakagami et al.~\cite{sakagami_2002}
achieved
$14.9\hbox{ Tflops}/512\hbox{ PN}$ with
$1.7\times 10^{10}$ grid points
for a fluid simulation.
(They used HPF to generate flat MPI processes.)
Komatitsch et al.~achieved
$5\hbox{ Tflops}/243\hbox{ PN}$
with $5.5\times 10^9$ grid points~\cite{komatitsch_2003}
for seismic wave propagation.
These results clealy indicate the high potential
of the Yin-Yang grid as a base grid system
on massively parallel computers.

\begin{table}
\begin{tabular}{|r|r|r|r|r||r|}
\hline
Paper & Shingu\cite{shingu_2002}
       & Yokokawa\cite{yokokawa_2002}
       & Sakagami\cite{sakagami_2002}
       & Komatitsch\cite{komatitsch_2003}
       & Kageyama et al. \\ \hline
Flops/PN & 26.6T/640
          & 16.4T/512
          & 14.9T/512
          & 5T/243
          & 15.2T/512 \\ \hline
efficiency &
    65\% &
    50\% &
    45\% &
    32\% &
    46\% \\ \hline
grid points (g.p.)&
    $7.1\times 10^8$ &
    $8.6\times 10^9$ &
    $1.7\times 10^{10}$ &
    $5.5\times 10^9$ &
    $8.1\times 10^8$ \\ \hline
g.p./AP &
    $1.4\times 10^5$ &
    $2.1\times 10^6$ &
    $4.2\times 10^6$ &
    $2.8\times 10^6$ &
    $2.1\times 10^5$ \\ \hline
Flops/g.p. &
    38K &
    19K &
    0.87K &
    0.91K &
    19K \\ \hline
Simulation kind &
    fluid &
    fluid &
    fluid &
    wave propagation  &
    fluid \\ \hline
Field &
    atmosphere &
    turbulence &
    inertial fusion &
    seismic wave &
    geodynamo \\ \hline
Method &
    spectral &
    spectral &
    finite volume &
    spectral element &
    finite difference \\ \hline
Parallelization &
    MPI-microtask &
    MPI-microtask &
    HPF (flat MPI)&
    flat MPI &
    flat MPI \\ \hline
\end{tabular}
\label{tab:03}
\caption{Performances on the Earth Simulator reported at SC}
\end{table}

\section{Simulation Results}
For geodynamo study, it is necessary
to follow the time development of the MHD system
until the thermal convection flow and the
dynamo-generated magnetic field are both sufficiently developed
and saturated.
(Initially,
both the convection energy
and the magnetic energy are negligibly small.)
For the case of grid size of
$255\times 514\times 1538\times 2=4\times 10^8$
with $3888$ processes,
it took six hours of wall clock time
until both the dynamo-generated magnetic field
and convection flow energy
reached to a saturated, and balanced, level.
In normalized time units,
this is about $0.3\%$ of the magnetic free decay time.
This suggests that we need to integrate more time
until we observe the dynamical features of the geodynamo
such as the repeated dipole reversals \cite{ochi_1999,kageyama_1999,li_2002}.

\noindent
\begin{figure}[t]
  \begin{center}
   \includegraphics[width=0.85\textwidth]{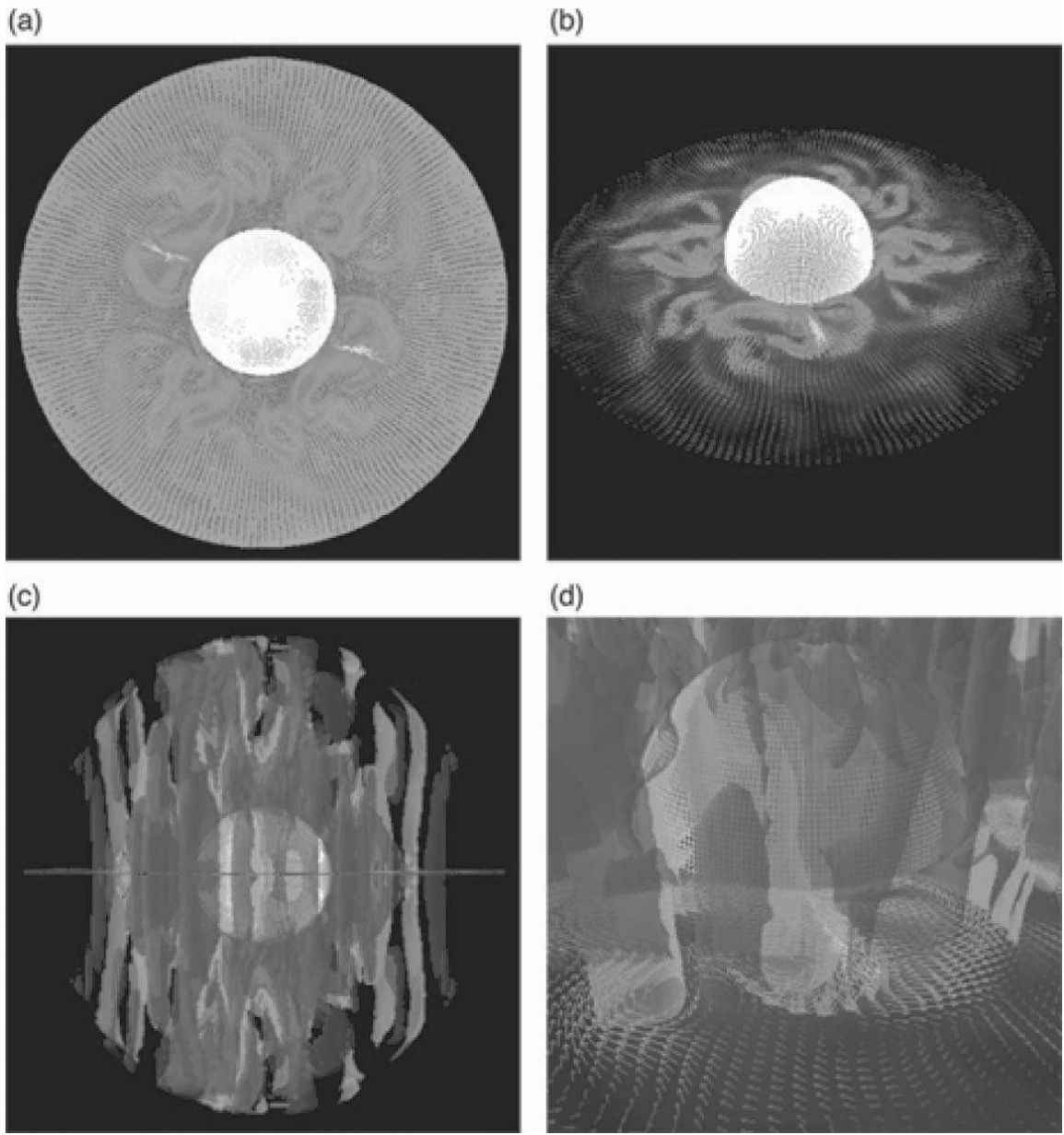}
  \end{center}
  \caption{Thermal convection structure obtained
by 3888 processor calculation on the Earth Simulator.
The total grid size is
$255 (\hbox{radial})
\times 514 (\hbox{latitudinal})
\times 1538 (\hbox{longitudinal})
\times 2 (\hbox{Yin and Yang})$.
In these visualizations,
grid points are reduced by a factor of $1/100$.
(a) Convection flow in the equatorial plane viewed from the north;
(b) Same data as (a) but viewed from $45^{\circ}N$;
(c) Columnar convection cells viewed in the equatorial plane.
Two colors indicate cyclonic and anti-cyclonic convection columns;
(d) Same data as (c), closer view.
}
  \label{fig:columns}
\end{figure}
As we described in Section~III,
the fundamental variables in our simulation are
the magnetic vector potential $\mathbf{A}=\{A_r,A_\theta,A_\phi\}$,
the mass flux density $\mathbf{f}=\{f_r,f_\theta,f_\phi\}$,
pressure $p$, and the mass density $\rho$.
It is convenient for data visualization/analysis purpose
to store the Cartesian components of
the magnetic field $\mathbf{B}=\{B_x,B_y,B_z\}$,
velocity $\mathbf{v}=\{v_x,v_y,v_z\}$,
vorticity  $\mathbf{\omega}=\{\omega_x,\omega_y,\omega_z\}$,
and temperature $T$.
During one simulation run of 6 hours of wall clock time,
we saved the 3-dimensional data 127 times, and
about $500$ GB of data was generated in total.

It is known that thermal convection motion
in a rapidly rotating spherical shell is organized
as a set of columnar convection cells.
The number of convection columns increases and
the flow and the generated magnetic field becomes
turbulent as shown in
Fig.~\ref{fig:columns},
which was obtained in the simulation of $3888$ processors
with $255\times 514\times 1538\times 2$ grid points.

\section{Summary}
In this paper, we reported the development
of a new geodynamo simulation code and presented
its performance on the Earth Simulator.
The base grid is the Yin-Yang grid,
which is a spherical overset grid composed of two identical
component grids.
The achieved performance is 15.2 Tflops by 4096 processors
with $511\times 514\times 1538\times 2$ grid points
in a spherical shell.

\begin{acknowledgments}
We would like to thank
Prof.~Tetsuya Sato, director of the Earth Simulator Center,
for his support and encouragement.
We also thank all members of the Earth Simulator Center,
especially Hiromitsu Fuchigami,
Shigemune Kitawaki,
Hitoshi Murai,
Satoru Okura,
and
Hitoshi Uehara
for their technical support and discussion.
We also thank Dr.~Daniel S.~Katz for polishing the
language in the manuscript.
\end{acknowledgments}


\begin{thebibliography}{10}

\bibitem{chesshire_1990}
G.~Chesshire and W.~D. Henshaw.
\newblock Composite overlapping meshes for the solution of partial differential
  equations.
\newblock {\em J. Comput. Phys.}, 90:1--64, 1990.

\bibitem{habata_2003}
Shinichi Habata, Mitsuo Yokokawa, and Shigemune Kitawaki.
\newblock {T}he {E}arth {S}imulator {S}ystem.
\newblock {\em NEC Res. \& Develop.}, 44(1):21--26, 2003.

\bibitem{hirai_2004}
Keisuke Hirai, Keiko Takahashi, Hidenori Aiki, Koji Goto, and Kunihiko
  Watanabe.
\newblock Development of non-hydrostatic ocean circulation simulation code on
  the Earth Simulator.
\newblock {\em Proc. of The 2004 workshop on the solution of partial
  differential equations on the sphere}, page~65, 2004.

\bibitem{kageyama_1995}
A.~Kageyama, T.~Sato, K.~Watanabe, R.~Horiuchi, T.~Hayashi, Y.~Todo, T.H.
  Watanabe, and H.~Takamaru.
\newblock Computer simulation of a magnetohydrodynamic dynamo. {II}.
\newblock {\em Phys. Plasmas}, 2:1421--1431, 1995.

\bibitem{kageyama_1999}
Akira Kageyama, Marcia~M. Ochi, and Tetsuya Sato.
\newblock Flip-flop transitions of the magnetic intensity and polarity
  reversals in the magnetohydrodynamic dynamo.
\newblock {\em Phy. Rev. Lett.}, 82:5409--5412, 1999.

\bibitem{kageyama_1997}
Akira Kageyama and Tetsuya Sato.
\newblock Generation mechanism of a dipole field by a magnetohydrodynamic
  dynamo.
\newblock {\em Phys. Rev. E}, 55:4617--4626, 1997.

\bibitem{kageyama_2004}
Akira Kageyama and Tetsuya Sato.
\newblock The ``{Y}in-{Y}ang {G}rid'': An overset grid in spherical geometry.
\newblock {\em Geochem. Geophys. Geosys., in press; preprint:
  arXive:physics/0403123}, 2004.

\bibitem{komatitsch_2003}
Dimitri Komatitsch, Seiji Tsuboi, Chen Ji, and Jeroen Tromp.
\newblock A 14.6 billion degrees of freedom, 5 teraflops, 2.5 terabyte
  earthquake simulation on the {E}arth {S}imulator.
\newblock {\em Proceedings of the {ACM/IEEE} Supercomputing SC'2003
  conference}, 2003, 2003.

\bibitem{komine_2004}
Kenji Komine, Keiko Takahashi, and Kunihiko Watanabe.
\newblock Development of a global non-hydrostatic simulation code using
  yin-yang grid system.
\newblock {\em Proc. of The 2004 workshop on the solution of partial
  differential equations on the sphere}, pages 67--69, 2004.

\bibitem{kono_2002}
Masaru Kono and Paul~H. Roberts.
\newblock Recent geodynamo simulations and observations of the geomagnetic
  field.
\newblock {\em Rev. Geophys.}, 40:1013, 2002.

\bibitem{li_2002}
Jinghong Li, Tetsuya Sato, and Akira Kageyama.
\newblock Repeated and sudden reversals of the dipole field generated by a
  spherical dynamo action.
\newblock {\em Science}, 295:1887--1890, 2002.

\bibitem{nakajima_2002}
Kengo Nakajima.
\newblock Three-level hybrid vs. flat {MPI} on the {E}arth {S}imulator:
  Parallel iterative solvers for unstructured grids on {GeoFEM} platform.
\newblock {\em RIST/TOKYO GeoFEM Report}, 2002-007, 2002.

\bibitem{ochi_1999}
Marcia~M. Ochi, Akira Kageyama, and Tetsuya Sato.
\newblock Dipole and octapole field reversals in a rotating spherical shell:
  Magnetohydrodynamic dynamo simulation.
\newblock {\em Physics of Plasmas}, 6:777--787, 1999.

\bibitem{ohdaira_2004}
Mitsuru Ohdaira, Keiko Takahashi, and Kunihiko Watanabe.
\newblock Validation for the solution of shallow water equations in spherical
  geometry with overset grid system.
\newblock {\em Proc. of The 2004 workshop on the solution of partial
  differential equations on the sphere}, page~71, 2004.

\bibitem{sakagami_2002}
Hitoshi Sakagami, Hitoshi Murai, Yoshiki Seo, and Mitsuo Yokokawa.
\newblock 14.9 {TFLOPS} three-dimensional fluid simulation for fusion science
  with HPF on the {E}arth {S}imulator.
\newblock {\em Proceedings of the {ACM/IEEE} {S}upercomputing {SC}'2002
  conference}, 2002.

\bibitem{shingu_2002}
Satoru Shingu, Hiroshi Takahara, Hiromitsu Fuchigami, Masayuki Yamada,
  Yoshinori Tsuda, Wataru Ohfuchi, Yuji Sasaki, Kazuo Kobayashi, Takashi
  Hagiwara, Shin ichi Habata, Mitsuo Yokokawa, Hiroyuki Itoh, and Kiyoshi
  Otsuka.
\newblock A 26.58 {T}flops global atmospheric simulation with the spectral
  transform method on the {E}arth {S}imulator.
\newblock {\em Proceedings of the ACM/IEEE Supercomputing SC'2002 conference},
  2002.

\bibitem{steger_1983}
Joseph~L. Steger, F.~Carroll Dougherty, and John~A. Benek.
\newblock A {C}himera grid scheme.
\newblock {\em Advances in Grid Generation}, edited by K.N. Ghia and U.
  Ghia:59--69, 1983.

\bibitem{takahashi_2004}
Keiko Takahashi et~al.
\newblock Development of nonhydrostatic coupled ocean-atmosphere simulation
  code on the earth simulator.
\newblock {\em Proc.of 7th International Conference on High Performance
  Computing and Grid in Asia Pacific Region}, Omiya, Japan, 2004.

\bibitem{xindong_2004}
Peng Xindong, Feng Xiao, Keiko Takahashi, and Takashi Yabe.
\newblock Conservative CIP transport in meteorological model.
\newblock {\em JSME International Journal}, {to be published}, 2004.

\bibitem{yokokawa_2002}
Mitsuo Yokokawa, Ken'ichi Itakura, Atsuya Uno, Takashi Ishihara, and Yukio
  Kaneda.
\newblock 16.4-tflops direct numerical simulation of turbulence by a Fourier
  spectral method on the earth simulator.
\newblock {\em Proceedings of the ACM/IEEE Supercomputing SC'2002 conference},
  2002.

\bibitem{yoshida_2004}
Masaki Yoshida and Akira Kageyama.
\newblock Application of the {Y}in-{Y}ang grid to a thermal convection of a
  {B}oussinesq fluid with infinite {P}randtl number in a three-dimensional
  spherical shell.
\newblock {\em Geophys. Res. Lett.}, 31:doi:10.1029/2004GL019970, 2004.

\end{thebibliography}

\end{document}